\documentclass[twocolumn, twocolappendix]{aastex701}
\usepackage{amsmath}

\begin{document}

\title{Revised AGN Spectral Model Reveals a More Significant Role in Cosmic Reionization}

\author[0009-0000-3702-1954]{Tong Su}
\affiliation{Institute for Frontiers in Astronomy and Astrophysics, Beijing Normal University, Beijing 102206, People’s Republic of China}
\affiliation{School of Physics and Astronomy, Beijing Normal University, Beijing 100875, People’s Republic of China}
\affiliation{Key Laboratory for Computational Astrophysics, National Astronomical Observatories, Chinese Academy of Sciences, Beijing 100012, China}
\affiliation{School of Astronomy and Space Sciences, University of Chinese Academy of Sciences, 19A Yuquan Road, Beijing 100049, China}
\email{sutong@bao.ac.cn}

\author[0000-0002-7972-3310]{Qi Guo}
\affiliation{Institute for Frontiers in Astronomy and Astrophysics, Beijing Normal University, Beijing 102206, People’s Republic of China}
\affiliation{School of Physics and Astronomy, Beijing Normal University, Beijing 100875, People’s Republic of China}
\affiliation{Key Laboratory for Computational Astrophysics, National Astronomical Observatories, Chinese Academy of Sciences, Beijing 100012, China}
\affiliation{School of Astronomy and Space Sciences, University of Chinese Academy of Sciences, 19A Yuquan Road, Beijing 100049, China}
\email[show]{qguo@bnu.edu.cn}

\author[0009-0000-0690-5562]{Wenxiang Pei}
\affiliation{Shanghai Key Lab for Astrophysics, Shanghai Normal University, Shanghai 200234, People’s Republic of China}
\email{wxpei@shnu.edu.cn}

\author[0000-0003-4176-6486]{Linhua Jiang}
\affiliation{Kavli Institute for Astronomy and Astrophysics, Peking University, Beijing 100871, China}
\affiliation{Department of Astronomy, School of Physics, Peking University, Beijing 100871, China}
\email{}

\begin{abstract}

Reionization marks one of the most important phase transitions in the history of the Universe, during which neutral baryonic matter was transformed into ionized plasma. While star-forming galaxies are widely regarded as the primary drivers of this process, the extent to which active galactic nuclei (AGNs) contribute remains a subject of ongoing investigation. In this study, we integrate a physically motivated AGN spectral energy distribution (SED) model with state-of-the-art observations to reassess the contribution of AGNs to cosmic reionization. Our findings indicate that adopting a more sophisticated AGN SED model could substantially increase the predicted ionizing photon output by a factor of 3$\sim$4, elevating AGNs to a more significant role ($\approx$20\%) in maintaining reionization than previously estimated. The inclusion of abundant faint AGNs further amplifies this contribution by a factor of a few. These conclusions remain robust across a wide range of accretion rates and ionizing photon escape fractions. Collectively, our results suggest that AGNs may have played a more prominent and previously underestimated role in the reionization of the Universe.

\end{abstract}



\section{Introduction} 

\setcounter{footnote}{0}   
The Epoch of Reionization (EoR) marks a critical phase in cosmic history when the intergalactic medium transitioned from neutral to ionized, ending the cosmic "dark ages" and rendering the Universe transparent to ultraviolet (UV) light. This transformation was driven by the first luminous sources - stars, galaxies, and possibly active galactic nuclei (AGNs). Despite extensive efforts, the dominant sources of ionizing photons remain uncertain \citep{Volonteri2009, Madau2015, Hassan2018, Finkelstein2019, 2023MNRAS.518.3576T}.
While star-forming galaxies are widely considered to be the primary drivers \citep{Bouwens2012, Finkelstein2012, Duncan2015, Robertson2015}, the contribution of active galactic nuclei (AGN) has been a subject of considerable debate. 

Early studies argued that the steep decline in AGN number density beyond redshift $z=4$ renders them insufficient to account for hydrogen reionization \citep{Shapiro1987, Shapiro1994, Giroux1996, Madau1999}. Recently, \citet{Giallongo2015, Giallongo2019} discovered a rich population of faint AGNs in the CANDELS fields at $4<z<6$, with number density reaching $\Phi\approx10^{-5}\mathrm{Mpc^{-3}mag^{-1}}$ for $M_\mathrm{UV}=-19$ at $z\approx6$. \citet{Onoue2017} compiled quasar luminosity function (QLF) measurements, showing $<10^{-5}\,\mathrm{Mpc^{-3}mag^{-1}}$ number density at $M_{1450}=-18$. By integrating the QLF down to $M_\mathrm{1450\text{\AA}}=-18$ and using the broken power law for the rest-frame quasar SED \citep{Lusso2015}, the contribution of quasar to the ionising photon budget at $z\approx 6$ is found to be minor. 
However, new observations with the James Webb Space Telescope \citep[JWST;][]{2006SSRv..123..485G} have uncovered a larger population of faint AGN candidates, reigniting interest in their potential role. An intriguing population of so-called “Little Red Dots” (LRDs; \citealt{Matthee2024}) at $z = 4-6$ has attracted significant attention through deep JWST/NIRCam imaging and wide-field slitless spectroscopy from the EIGER and FRESCO surveys.
\citet{Harikane2023} identified a high number density of type-1 AGN using JWST/NIRSpec deep spectroscopy, with $\Phi\approx10^{-4} \mathrm{Mpc^{-3}mag^{-1}}$ for $M_\mathrm{UV}\approx-18.5$ at z=6. The type-2 AGN survey JADES \citep{Maiolino2024, Scholtz2025} conservatively suggests a type-2-to-type-1 ratio of two at $z\sim4-6$, rendering the type-2 AGN number density to be almost two orders of magnitude higher than the X-ray selected AGN number density reported by \citet{Giallongo2019}.
Using recent data, \citet{2024ApJ...971...75M} suggests that AGN alone could drive reionization, provided a high fraction of broad-line AGN among galaxies and large hydrogen escape fractions.

Moreover, previous estimates often relied on simplified conversions from observed UV luminosities to total ionizing photon output, potentially underestimating or overestimating AGN contributions. 
It is important to note that the contribution of AGN to the ionizing photon budget depends sensitively on the adopted AGN spectra energy distribution. 
\citet{Yoshiura2017} considered two SED models - an AC model (accretion disc + corona) and a PL model (broken power-law SED by \citet{Lusso2015}). The AC model produces significantly more ionizing photons than the PL model, leading to a more efficient HI ionization, while the PL model is ruled out for AGN-driven reionization.

In this work, we incorporate a physically motivated AGN SED model \citep{Su2025} along with state-of-the-art observations. By considering different scenarios on the black hole (BH) accretion rate and the photon escape fraction, we aim to provide new insights into the role of AGN in cosmic reionization. This paper is structured as follows: Sec.\,\ref{sec:data} introduces the dataset used in this work; Sec.\,\ref{sec:models} describes the two SED models; Sec.\,\ref{sec:methods} describes our methods and their differences with those of previous works; We present and summarize the main results in Sec.\,\ref{sec:results} and Sec.\,\ref{sec:summary}.

\section{Data and Methods}
\subsection{AGN luminosity functions}\label{sec:data}
We compile UV (1450\AA) luminosity functions (LFs) from both recent JWST-based studies and earlier pre-JWST results, including measurements by \citet{Jiang2022} at $z\approx6.2$, \citet{Kokorev2024} at $6.5<z<8.5$, and \citet{Harikane2023} at $z=5.76$ and $z=6.94$.
\citet{Jiang2022} derived their cumulative LF from a combination of the Subaru High-z Exploration of Low-Luminosity Quasars (SHELLQs) project observation \cite{Matsuoka2018} and the upper limit of four zero-detection fields. Their number density at \(M_{\rm UV} = -18\) is \(\Phi \approx \mathrm{10^{-4.7}\,Mpc^{-3}\,mag^{-1}}\) at $z=6.2$.
In contrast, \citet{Harikane2023} identified a sample of faint type-1 AGNs using deep JWST/NIRSpec spectroscopy and reported a faint-end number density roughly an order of magnitude higher than that in \citet{Jiang2022}. 
For comparison, we also include the Little Red Dots in deep JWST/NIRCam fields. \citet{Kokorev2024} found that their UVLF at $6.5 < z < 8.5$ exhibits a similar faint-end number density to that of \citet{Jiang2022}. 
A compilation of these UVLF measurements is presented in Appendix.\,\ref{append:UVLF}.

\subsection{The broken power-law and disk-corona model}\label{sec:models}
In this work, the ionizing photon production efficiency $\xi(M_\mathrm{UV})$ is derived from the disk-corona model \citep{Su2025} as a function of $M_{1450}$, in the contrary to many previous studies that adopt a broken power-law SED model \citep{Lusso2015} that peaks at 912\AA~ to estimate $\xi(M_\mathrm{UV})$.

The broken power-law model in \cite{Lusso2015} has the form $f_\nu\propto\nu^{\alpha_\nu}$, where $\alpha_\nu=-0.61$ at $\lambda>912$\AA~ and $\alpha_\nu=-1.7$ at $\lambda<912$\AA. This spectrum is obtained by stacking 53 luminous quasar spectra at $z\approx2.4$ across rest-frame wavelength 600-2500\AA, it is therefore constrained to that limited spectral interval. Although widely used to assess the contribution of AGN to the epoch of reionization \citep{Madau2015, Hassan2018, Onoue2017, Jiang2022}, its physical applicability is limited by the systematic differences in typical black hole accretion rates between the EoR and cosmic noon. 
Moreover, the spectral break is fixed at $912\,$\AA~ by construction, while both observational and theoretical studies suggest that the peak frequency of AGN SEDs can vary significantly with black hole mass and accretion rate \citep{Shakura1973, Done2012, 2015MNRAS.449.2174C, Thomas2016}, and that the high-frequency behavior of the SED may deviate from a simple broken power law \citep{Magdziarz1995, Dadina2008, Done2012}.

In the magnetic reconnection-heated disk-corona model developed by \citet{LiuBF2002}, the accretion flow consists of an underlying cold thin disk and hot coronae that sandwich it.
The corona is built up through energy equilibria between thermal conduction and mass evaporation, magnetic reconnection heating and Compton scattering cooling. By solving the energy balance equations globally, the coronal strength increases with decreasing radial distance to the central BH. The soft seed photon field emitted by the underlying disk undergoes multiple inverse-Compton scattering in the hovering hot corona; its energy is therefore shifted higher, contributing to high-energy emissions. 
In the \citet{Su2025} model, the energy equilibrium equations are solved for each annulus individually, rather than globally as in the original model. The modified model further considers super-Eddington accretion. As the BH accretion rate increases, an optically thick, radiative pressure-dominated, slim disk-like region would emerge in the inner part of the disk, replacing the disk-corona configuration. This part of the accretion flow ceases to produce high-energy photons, leading to a softer total spectrum as the accretion rate increases, aligning with observational evidence \citep{2015MNRAS.447.1692Y, 2016MNRAS.459.3963C}.
This model has been shown to reproduce both the SEDs of individual local AGNs and the AGN luminosity functions at redshifts $z < 1$.

\begin{figure*}
\centering
{\includegraphics[width=\columnwidth]{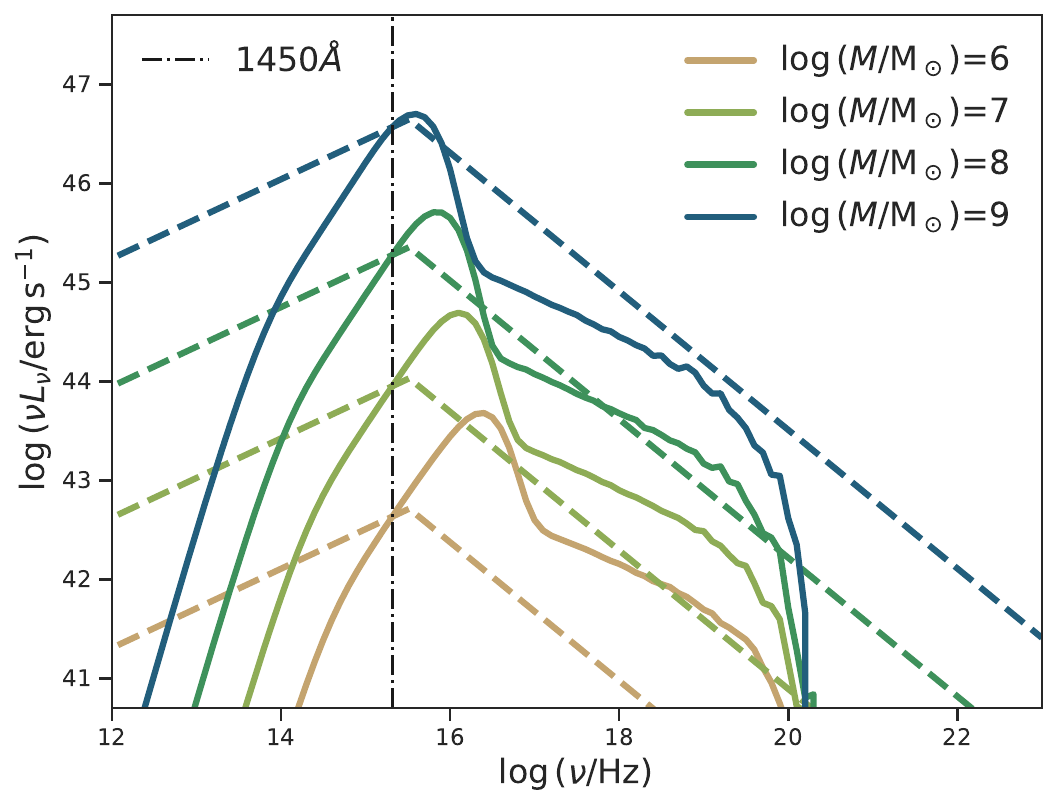}}
{\includegraphics[width=\columnwidth]{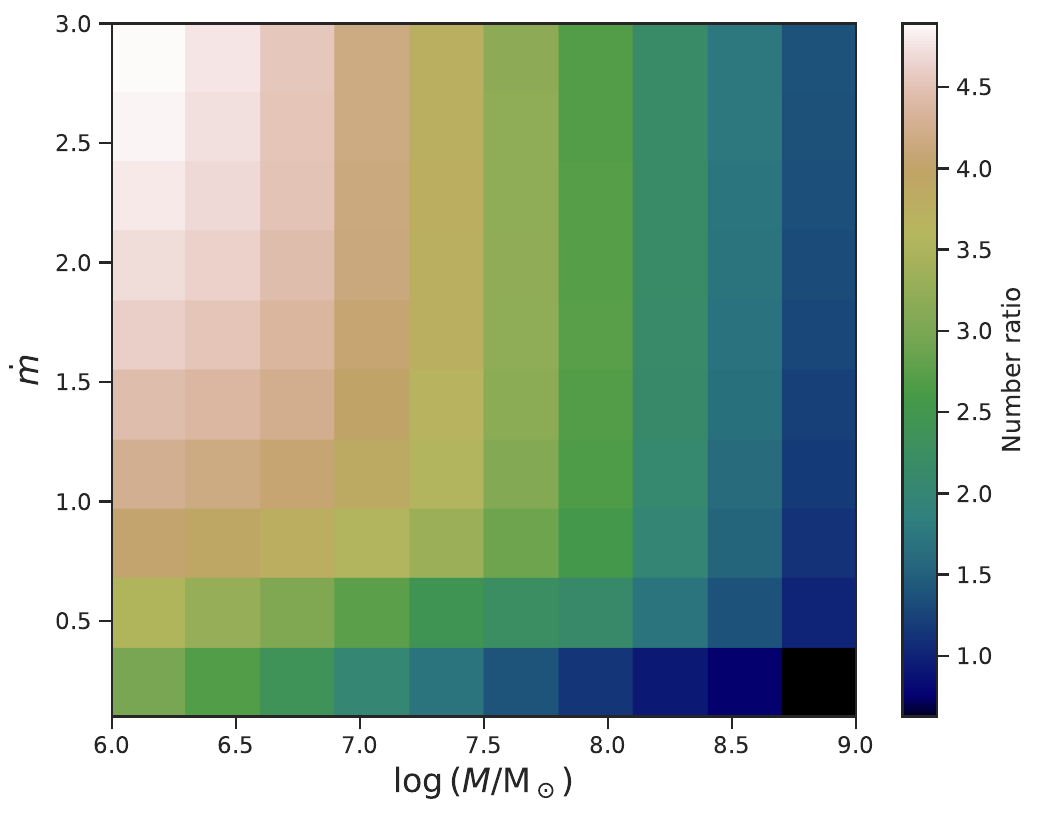}}
\caption{
{\bf SED comparison and ionizing photon number ratio.}
\textit{Left panel:} Spectral energy distributions (SEDs) of the broken power-law model \citep{Lusso2015} (dashed curves) and the disk-corona model \citep{Su2025} (solid curves), color-coded by black hole mass. An Eddington accretion rate is assumed for the disk-corona model. The broken power-law SEDs are normalized to match the UV magnitude ($M_{1450}$) of the corresponding disk-corona cases. The vertical black dash-dotted line marks 1450\AA. 
\textit{Right panel:} Ionizing photon number ratio between the two models, $\dot{n}_\mathrm{ion,diskcor}/\dot{n}_\mathrm{ion,PL}$, indicated by the color scale. The peak frequency of the disk-corona model shifts away from 1450\,\AA~with decreasing black hole mass and increasing accretion rate. The ratio reaches its maximum at the lowest black hole mass and the highest accretion rate. 
} 
\label{fig:SEDs}
\end{figure*}

\subsection{Ionizing photon production efficiency}\label{sec:methods}
The left panel Fig.\,\ref{fig:SEDs} compares the SEDs predicted by the disk-corona model \citep{Su2025} and the broken power-law model \citep{Lusso2015}, assuming Eddington-limited accretion (see Appendix \,\ref{append:SEDs} for a detailed comparison between the model SEDs and observations.). The primary difference lies in the spectral shape and the location of the peak frequency. In the power-law model, both remain fixed regardless of black hole properties. In contrast, the disk-corona model predicts a peak frequency that increases with decreasing black hole mass, resulting in SEDs that can extend well beyond the Lyman limit (912\AA) for lower-mass AGNs. This leads to substantial differences in the predicted ionizing photon production rate, with important implications for estimating AGN contributions to cosmic reionization.

A more detailed comparison of the ionizing photon production efficiency between the disk-corona and power-law models is presented in the right panel of Fig.\,\ref{fig:SEDs}, which shows the ratio $\xi_\mathrm{ion,diskcor}/\xi_\mathrm{ion,PL}$ as a function of BH mass and accretion rate. This ratio peaks for low-mass supermassive black holes (SMBHs) accreting at near- or super-Eddington rates, where the spectral peak in the disk-corona model shifts into the extreme ultraviolet, substantially boosting the ionizing photon output. As the black hole mass increases or the accretion rate decreases, the spectral peak moves to longer wavelengths, leading to a decline in ionizing efficiency relative to the power-law model. This behavior reflects the intrinsic dependence of the disk-corona SED on black hole properties, in contrast to the fixed spectral shape assumed in the broken power-law prescription. The faint AGNs detected by JWST observations are likely powered by SMBHs with masses of $\approx 10^6–10^8\,\mathrm{M_\odot}$ and accreting at or above the Eddington limit \citet{Harikane2023}.
Within this parameter space, the disk-corona model predicts a typical ionizing efficiency enhancement of $\xi_\mathrm{ion,diskcor}/\xi_\mathrm{ion,PL} \approx 2-4$, highlighting the importance of physically motivated SED models in accurately assessing the contribution of faint AGNs to cosmic reionization.

\section{Results}\label{sec:results}
In this section, we investigate the AGN contribution in cosmic reionization using the modified disk–corona model described above. We also examine the impact of varying escape fractions and the growth history of faint AGN black holes.

\begin{figure}[h]
\centering
{\includegraphics[width=\columnwidth]{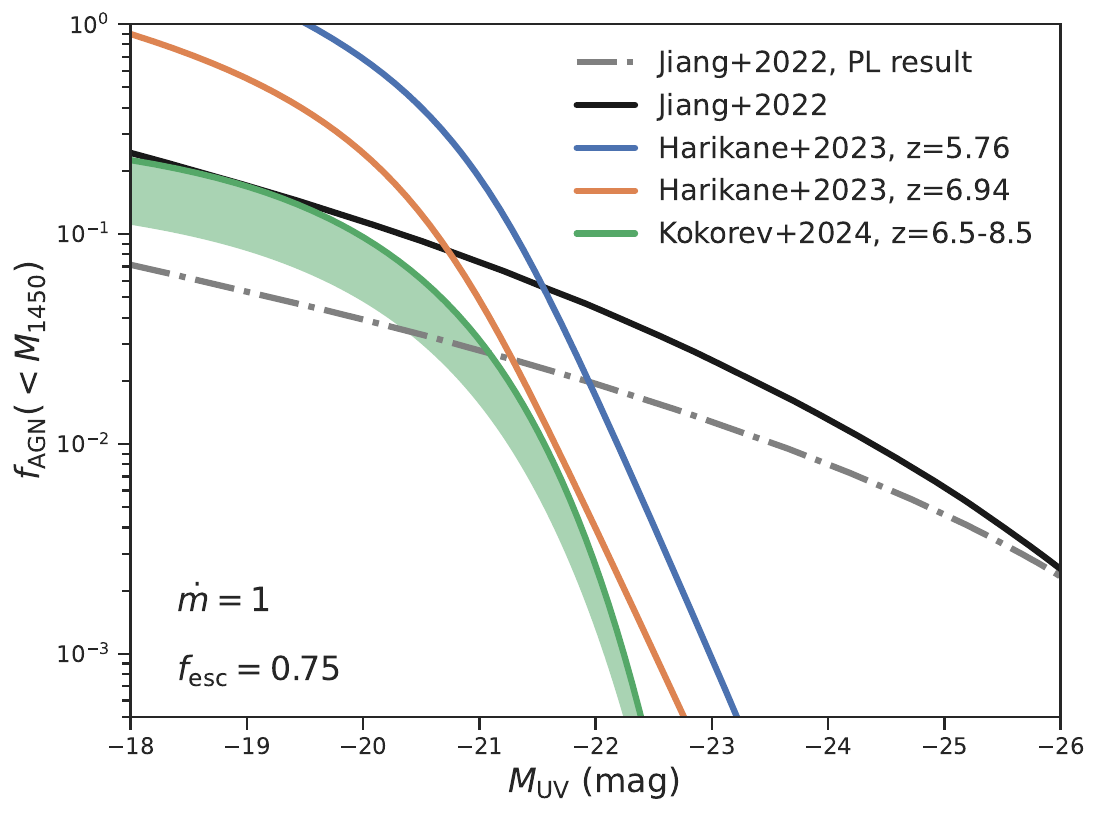}}
\caption{
{\bf The AGN contribution to reionization.}
The solid curves represent the AGN contributions to the ionizing photon budget, calculated using the disk-corona model \citep{Su2025}. An Eddington accretion rate is assumed in computing the disk-corona SEDs. The escape fraction adopted throughout this panel is $f_\mathrm{esc} = 0.75$. The blue and orange curves correspond to the AGN contributions based on the \citet{Harikane2023} luminosity functions at $z = 5.76$ and $z = 6.94$, respectively. The green shaded region indicates the range of contributions derived from the \citet{Kokorev2024} luminosity function across the redshift interval $6.5 < z < 8.5$. 
For comparison, the gray dash-dotted curve reproduces the result from \citet{Jiang2022}, which estimates an AGN contribution of approximately 7\% using the ionizing efficiency derived from integrating the broken power-law SED model \citep{Lusso2015}, also assuming $f_\mathrm{esc} = 0.75$.
}
\label{fig:fagn_Edd}
\end{figure}

We define the AGN contribution in maintaining cosmic reionization at given redshift as $f_\mathrm{AGN}(<M_{1450}) = \dot{N}_\mathrm{AGN}(z) / \dot{N}_\mathrm{ion}(z)$, 
where this ratio can be interpreted as the fraction of the ionizing photons produced by AGN relative to the total photon production rate required to balance recombinations in the intergalactic medium. In other words, maintaining cosmic reionization refers to providing enough ionizing photons to counteract the continuous recombination of ionized hydrogen to keep the Universe remain ionized at a certain redshift.
The comoving photon production rate of AGN is calculated as $\dot{N}_\mathrm{AGN}(z)=\int^{M_{1450}}\int_{\nu_{1\mathrm{Ryd}}}^{\nu_{4\mathrm{Ryd}}}\frac{L_\nu}{\mathrm{h}\nu}\mathrm{d}\nu \times \Phi(M_\mathrm{1450}, z)f_\mathrm{esc}\mathrm{d}M_{1450}$, where $L_\nu$ is the SED of an AGN with certain 1450\AA~absolute magnitude, $\Phi(M_\mathrm{1450}, z)$ is the AGN number density, and $f_\mathrm{esc}$ is the escape fraction.
The critical comoving ionizing photon production rate required to keep the Universe ionized at redshift $z$ is given by $\dot{N}_\mathrm{ion}(z) = 10^{50.48} \left( \frac{C}{3} \right) \left( \frac{1+z}{7} \right)^3\,\mathrm{Mpc^{-3}\,s^{-1}}$ \citep{Madau1999}, assuming case B recombination. Here, $C$ is the clumping factor, with values typically predicted by simulations to lie in the range $C \approx 2-5$ during reionization \citep{2009MNRAS.394.1812P, Pawlik2015}, we adopt $C = 3$ as a representative value. 
Given that most high-redshift AGNs are observed to accrete near the Eddington limit, we adopt Eddington-limited accretion as our fiducial assumption.

The main results are presented in Fig.\,\ref{fig:fagn_Edd}, which shows the predicted AGN contribution to maintain the hydrogen reionization based on the disk-corona SED model, assuming an ionizing photon escape fraction of $f_\mathrm{esc} = 0.75$ \citep{2016MNRAS.462.2478C}. 
Integrating the AGN luminosity functions from \citet{Jiang2022} to $M_{1450}=-18$, the disk-corona model predicts an AGN contribution of approximately 24\% to the required ionizing photon budget at $z = 6.2$. This is substantially higher than the $\approx7\%$ contribution estimated under the broken power-law model \citep[][the grey dashed curve]{Jiang2022}, highlighting that AGNs - particularly those with lower-mass black holes and high accretion rates - may represent a non-negligible component in the cosmic reionization budget. 
When combined with the AGN luminosity functions derived by \citet{Harikane2023} at $z = 5.76$ and $z = 6.94$, the predicted AGN contribution to the ionizing photon budget becomes even more significant. At $z = 5.76$, the cumulative contribution from AGNs - including faint sources with $-18>M_{1450} > -22$ - can exceed unity, suggesting that the Universe may have been fully ionized at an earlier epoch than conventionally assumed. Although LRDs may have substantially lower ionizing escape fractions than typical AGNs, we adopt the same value for comparison. Using the UV luminosity function (UVLF) of LRDs from \citet{Kokorev2024}, the resulting ionizing photon production is comparable to that derived from the luminosity function of \citet{Jiang2022}.

\begin{figure*}
\centering
\includegraphics[width=\columnwidth]{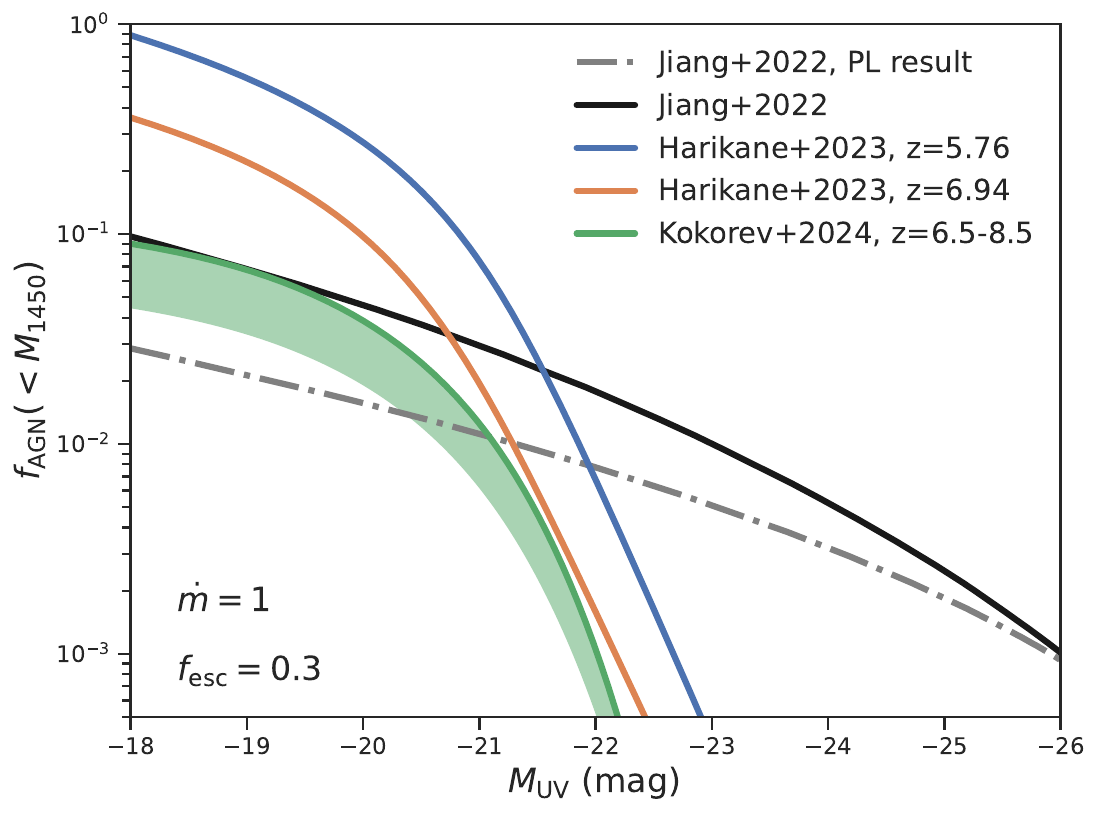}
\includegraphics[width=\columnwidth]{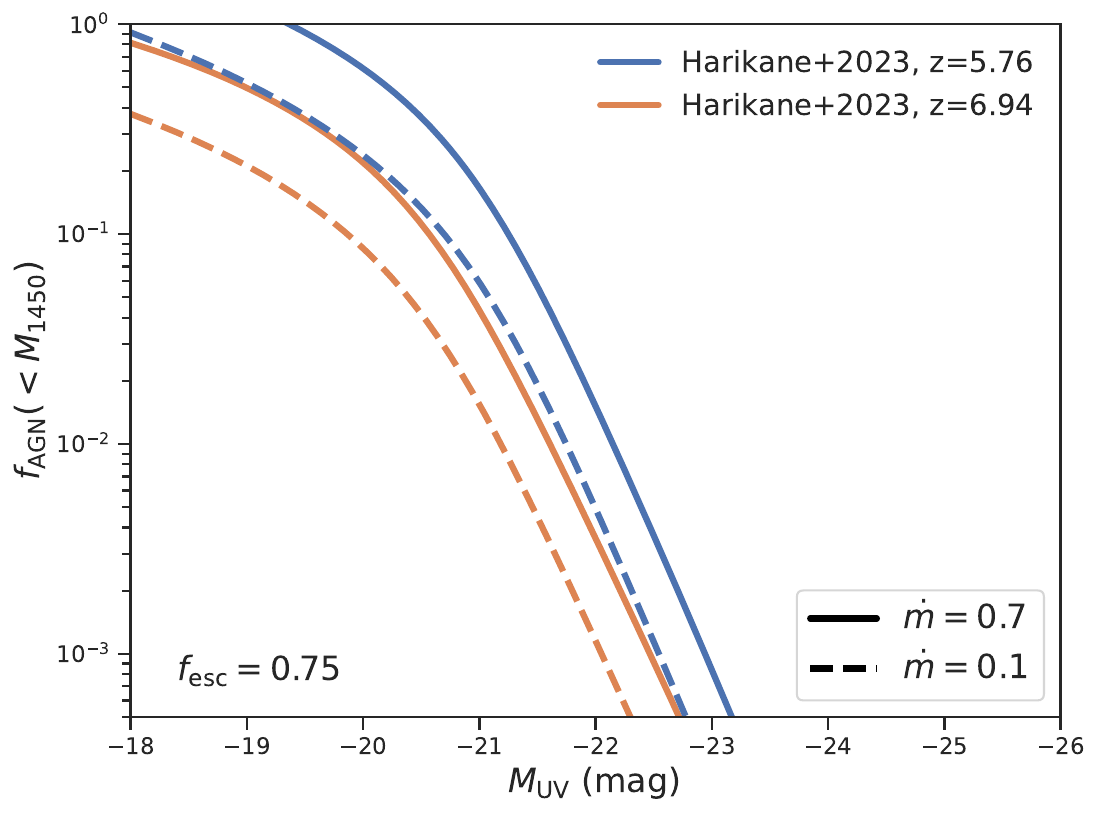}
\caption{
{\bf The AGN contribution to reionization: dependence on escape fraction and accretion rate.}
\textit{Left panel:} Same as Fig.\,\ref{fig:fagn_Edd}, but adopting a lower escape fraction of $f_\mathrm{esc} = 0.3$. Under this assumption, the type-1 AGN population reported in \citet{Harikane2023} still contributes approximately 36-88\% to the ionizing photon budget. In contrast, AGN contributions based on the \citet{Jiang2022} and \citet{Kokorev2024} luminosity functions are reduced to around 10\% due to their lower faint-end number densities. The broken power-law model, combined with the \citet{Jiang2022} luminosity function, predicts an even smaller contribution of roughly 3\%.
\textit{Right panel:} AGN contributions computed using the $z = 5.76$ and $z = 6.94$ luminosity functions from \citet{Harikane2023}, assuming escape fraction $f_\mathrm{esc} = 0.75$ and two different accretion rates: $\dot{m} = 0.7$ and $\dot{m} = 0.1$. Here, $\dot{m} = 0.7$ represents the lower limit of the Eddington-normalized accretion rate required to reach typical LRD properties, as discussed in the main text; $\dot{m} = 0.1$ corresponds to the lower bound of accretion rates inferred in \citet{Harikane2023}.
}
\label{fig:fagn_diff_scenario}
\end{figure*}

While an escape fraction of $f_\mathrm{esc} = 0.75$ is often adopted based on studies of low-redshift AGNs, its actual value at high redshift remains uncertain. To assess the sensitivity of our results to this parameter, we consider a more conservative scenario with $f_\mathrm{esc} = 0.3$. As shown in the right panel of Fig.\,\ref{fig:fagn_diff_scenario}, even under this reduced escape fraction, AGNs still contribute approximately 36-88\% of the required ionizing photon budget when using the luminosity functions from \citet{Harikane2023}.

Another key assumption underlying the preceding analysis is the Eddington limit accretion scenario. 
While this is largely supported by recent observations of high-redshift AGN \citep[e.g.,][]{Harikane2023, Kokorev2024, Maiolino2024}, 
it is plausible that some AGNs experienced more gradual growth in the early Universe. To explore this possibility, we consider a simplified model in which black holes originate from Population~III remnant BH seeds\footnote{There are three commonly discussed formation channels for SMBH seeds: (1) remnants of Population~III stars, (2) runaway collapses of dense stellar clusters, and (3) direct collapse of massive gas clouds. 
Given the number density of type~1 AGNs reported in recent JWST observations, $\Phi \approx \mathrm{10^{-4}\,Mpc^{-3}\,mag^{-1}}$ at the faint end, Population~III remnants appear to be the most plausible candidates, as the expected number densities for the other two channels are significantly lower.} 
We consider a scenario where a BH with a typical initial mass of {$100\,\mathrm{M_{\odot}}$ \citep[e.g.,][]{2002ApJ...571...30S, 2003ApJ...592..645Y, 2004ARA&A..42...79B}} seeded at redshift $z = 15$, corresponding to the expected onset of galaxy formation. We assume a constant accretion rate that is a fixed fraction of the Eddington limit. Under this framework, in order to reach a black hole mass of $\approx10^6\,\mathrm{M_{\odot}}$ and an absolute rest-frame UV magnitude of $M_\mathrm{1450} \approx -18$ (the typical lower bounds for high-redshift AGNs) by redshift $z = 6$, a minimum accretion rate of approximately 0.7 times the Eddington rate is required (see Appendix.\,\ref{append:LRD_growth} for details). The inferred accretion rate lies close to the Eddington limit, lending support to the our fiducial model assumption that early black holes grew in a near-Eddington accretion regime.
The left panel of Fig.\,\ref{fig:fagn_diff_scenario} utilizes this growth scenario, and combines it with the AGN luminosity functions from \citet{Harikane2023}. As such, the AGN contribution to the ionizing photon budget $f_\mathrm{AGN}$ is estimated to remain $\gtrsim80\%$ at both redshifts analyzed. 

For comparison, we also explore a more extreme scenario with a significantly lower accretion rate of 0.1 times the Eddington rate\footnote{
 An AGN with $M_{\mathrm{UV}} = -18$ at $z = 6$ would correspond to a black hole mass of $M_{\mathrm{BH},\,z=6} \approx 10^{7}\,\mathrm{M_{\odot}}$ when assuming an accretion rate of $\dot{m} = 0.1$. If this accretion rate is maintained throughout the black hole's growth, it would need to originate from a seed of $M_{\mathrm{seed}} \approx 10^{6.34}\,\mathrm{M_{\odot}}$ at $z = 15$. The comoving number density of such massive seeds is expected to be extremely low, on the order of $\sim10^{-9}-10^{-6}\,\mathrm{Mpc^{-3}}$ \citep[e.g.,][]{2016MNRAS.463..529H, 2016MNRAS.457.3356V}. This is significantly below the observed number density of faint type~1 AGNs at $z = 6$, making this growth scenario highly unlikely. We acknowledge, however, that a black hole could in principle begin with a smaller seed and undergo an early phase of rapid accretion before declining to $\dot m = 0.1$ by $z =6$; such a history cannot be ruled out.}
As denoted with the dashed curves in the right panel of Fig.\,\ref{fig:fagn_diff_scenario}, even under this low-accretion scenario, the AGN contribution remains non-negligible, ranging from approximately 37-91\% at redshifts $z = 6.94$ and $z = 5.76$, respectively. These results collectively demonstrate that AGNs could provide a substantial fraction of the ionizing background during reionization, even without requiring continuous near-Eddington growth.

High-energy photons emitted by AGNs can, in principle, travel larger distances and produce multiple ionizations before becoming fully thermalized (add ref here)\citep{Shull1985, Madau2017}. 
Consequently, the AGN contributions derived above are based solely on UV photons (1-4\,Ryd), representing conservative lower limits. To evaluate the impact of including high-energy photons, we extend our calculation to high-energy photons with energies between 4\,Ryd-10\,keV. When the energy dependence of the photoionization cross section is ignored, the ratio of the photoionization rate produced by high-energy photons to that produced by UV photons is enhanced by approximately 52\% at $M_\mathrm{UV}= -18$ in a highly ionized scenario, and this enhancement declines rapidly toward higher luminosities. However, once the strong photon-energy-dependent cross section is included, the enhancement is reduced to only a few percent. 
We also assess the contribution from secondary ionizations produced by high-energy photons and find it to be negligible compared to the primary ionizations from UV photons, amounting to less than 0.0006\% in a highly ionized scenario when the energy-dependent cross section is taken into account (See Appendix.\,\ref{append:Xray_secondary} for details). 

Overall, our analysis shows that AGNs can provide a substantial fraction of the ionizing photon budget at $z \sim 6$ under a wide range of plausible assumptions. Even when adopting conservative values for the ionizing escape fraction and allowing for sub-Eddington black hole growth, the predicted AGN contribution remains high, reaching tens of percent to unity depending on the luminosity function employed. The impact of high-energy–photon–driven ionizations provides a modest additional boost to the total ionizing output of AGNs. These results point to a more significant role for AGNs in sustaining the ionization state of the intergalactic medium during the late stages of reionization than previously assumed.

\section{Discussions and conclusions}\label{sec:summary}
Our study shows that adopting a physically motivated AGN spectral energy distribution (SED) model leads to a substantial increase in the predicted ionizing photon output, implying that AGNs may have played a far more important role in the cosmic reionization than previously assumed. Unlike simplified power-law prescriptions, the disk-corona model captures the dependence of the ionizing efficiency on black hole mass and accretion rate, naturally enhancing the contribution from faint, rapidly accreting AGNs.

The inclusion of the abundant faint AGNs recently uncovered by JWST further strengthens this picture. Their high number densities significantly elevate the total ionizing photon budget, in some cases approaching or even exceeding the critical level required to sustain reionization by redshifts $z \approx 6-7$. These results remain robust when varying key assumptions such as the escape fraction of ionizing photons and the accretion history of faint black holes, underscoring the robustness of the conclusions across a wide parameter space.

{


However, we note that the clumping factor can significantly affect the inferred ionizing photon budget. In the above analysis, we adopt $C = 3$, a fiducial value commonly used in reionization studies \citep[e.g.,][]{2015ApJ...811..140B, Yoshiura2017, Jiang2022, 2024arXiv240407250M}, which is supported by hydrodynamic simulations \citep{2009MNRAS.394.1812P, Pawlik2015}. 
Recently, however, \citet{2024arXiv240618186D} derived a much higher clumping factor of $C \approx 12$, based on the photoionization rate and photon mean free path inferred from Ly$\alpha$ forest measurements and stacked quasar spectra. 
A larger clumping factor implies a shorter recombination timescale and therefore a higher ionizing photon budget to maintain ionization equilibrium. Adopting $C = 12$ reduces the fraction of ionizing photon AGN could provide to sustain reionization, $f_{\mathrm{AGN}}$, by a factor of four. When combined with the UVLF from \citet{Harikane2023}, this corresponds to $f_{\mathrm{AGN}} \approx 55\%$ at $z = 5.76$ and $f_{\mathrm{AGN}}\approx 22\%$ at $z = 6.94$ with our fiducial assumptions.



Recent JWST observations have revealed a surprisingly high number density of galaxies at $z \gtrsim 6$ \citep[e.g.,][]{2023ApJS..265....5H, 2023ApJ...946L..13F, 2024ApJ...969L...2F}. 
Recently \citet{2024MNRAS.535L..37M} has shown that the ionizing photon budget from galaxies alone may be more than sufficient to fully reionize the Universe by $z \approx 6$, assuming high ionizing efficiencies and an enhanced abundance of early ($z > 9$) galaxies. 
However, as in our analysis, these studies also depend on observational constraints on the escape fraction, clumping factor, faint-end luminosity functions, etc. 
It is therefore likely that both galaxies and AGNs contributed jointly to cosmic reionization. 
A quantitative assessment of their relative roles will require tighter observational constraints on both populations.

}

Taken together, our findings suggest that AGNs may have been more efficient in producing ionizing photons and contributing to cosmic reionization than previously recognized.
This has important implications for interpreting the earliest phases of structure formation, and highlights the need for future observational programs to better constrain the demographics and physical properties of faint AGNs during the Epoch of Reionization.


\section*{Acknowledgement}
We thank Yi Mao for helpful discussions. 
This work is supported by the National Natural Science Foundation of China (NSFC) (No. 12425303), the National SKA Program of China (No. 2022SKA0110201), the NSFC (grant Nos. 12588202, 12033008),
National Key R\&D Program of China (Grants No. 2023YFA1607903), the CAS Project for Young Scientists in Basic Research grant No. YSBR-062, the K.C.Wong Education Foundation, and the Strategic Priority Research Program of the Chinese Academy of Sciences, Grant No.XDB0500203.
QG acknowledges the hospitality of the International Centre of Supernovae (ICESUN), Yunnan Key Laboratory at Yunnan Observatories Chinese Academy of Sciences, and European Union's HORIZON-MSCA-2021-SE-01 Research and Innovation programme under the Marie Sklodowska-Curie grant agreement number 101086388. 

\appendix

\section{SEDs of individual sources}\label{append:SEDs}
\begin{figure*}[h!]
\includegraphics[width=\textwidth]{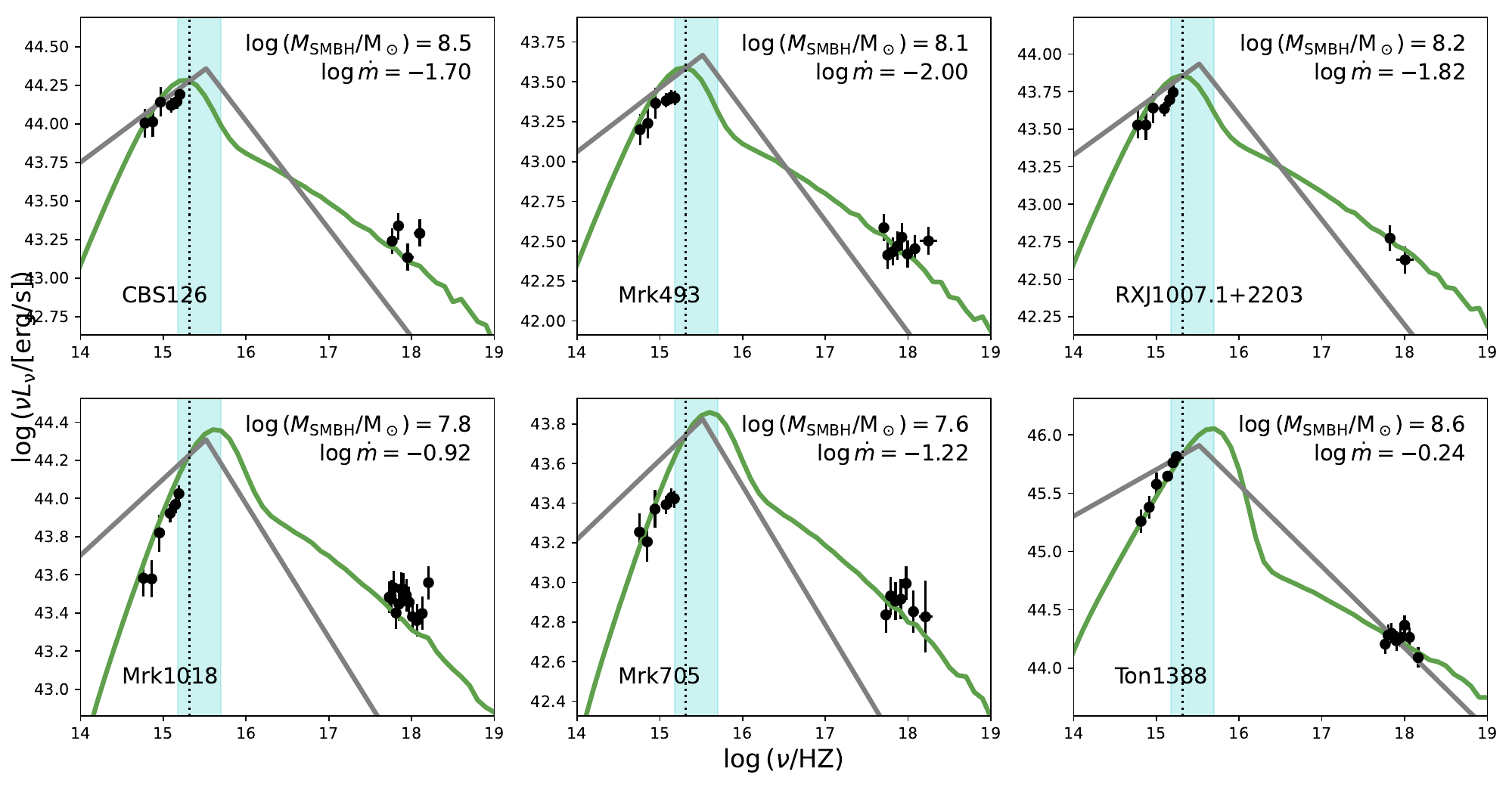}
\caption{ \textbf{Comparison between model-predicted and observed SEDs.
} 
Observed SEDs of CBS126, Mrk\,493, RXJ1007.1+2203, Mrk\,1018, Mrk\,705, and Ton\,1388 are shown as symbols with error bars, with data compiled from \citet{Cheng2019, Cheng2020}. Our disk-corona SED models, with the inferred SMBH mass and Eddington-normalized accretion rate indicated in each panel, are overplotted for comparison \cite[see][for model details]{Su2025}. The gray solid curves represent the broken power-law SED rescaled to match the UV luminosity at 1450\,\AA. The vertical black dotted line marks 1450\,\AA, and the cyan-shaded region highlights the wavelength range (2500-600\,\AA) over which the broken power-law slopes were defined \citep{Lusso2015}. Overall, the disk-corona models provide a closer match to the observed SED shapes than the broken power-law approximation.
\label{fig:source_diskcor}}
\end{figure*}

Fig.\,\ref{fig:source_diskcor} shows the observational data and our model-predicted SED for several local high-luminosity AGNs, overlapped by the broken power-law SED rescaled to the same UV luminosity. Our model SEDs are based on the disk-corona model, with the SMBH mass and accretion rate indicated in the upper-right corner of each panel. Observational data for CBS126, Mrk493, RXJ1007.1+2203, Mrk1018, Mrk705, and Ton1388 are taken from \citet{Cheng2020, Cheng2019}. 
In general, our model predictions are consistent with observational data and successfully reproduce the overall characteristics of observed AGN SEDs. However, in many cases the broken power-law model yields spectral slopes that differ from those inferred from the observations.

\section{UV luminosity functions}\label{append:UVLF}

\begin{figure*}
\centering
\includegraphics[width=\columnwidth]{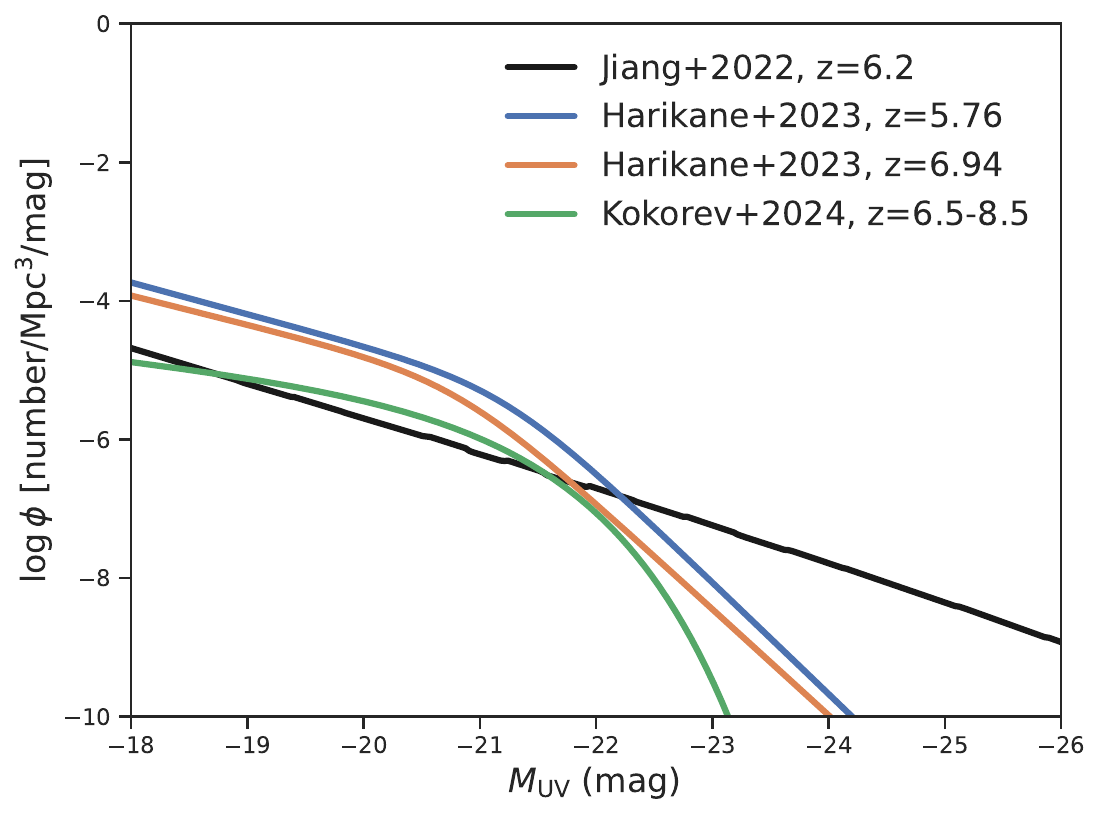}
\includegraphics[width=\columnwidth]{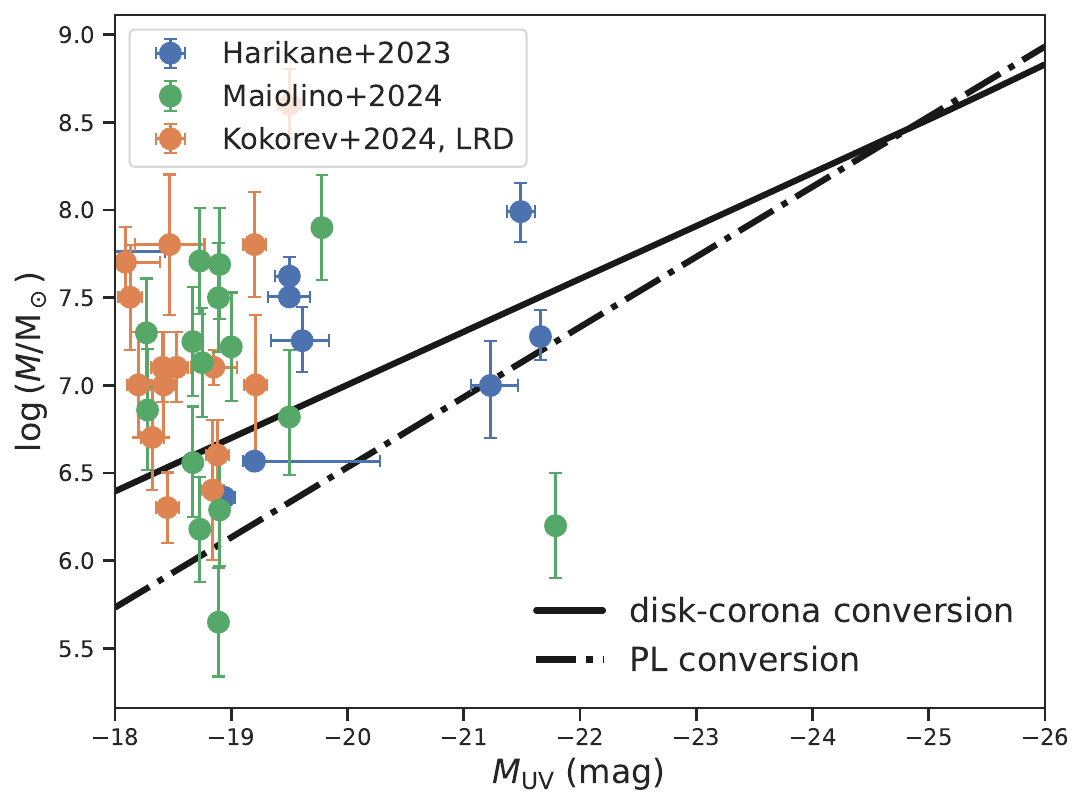}
\caption{
{\bf  UV luminosity functions and UV magnitude vs. SMBH mass relation.}
\textit{Left panel:} AGN UV luminosity functions compiled from previous observational studies \citep{Jiang2022, Harikane2023, Kokorev2024}. The black curve shows the pre-JWST constraints from \citet{Jiang2022} (95\% confidence level); the green curve corresponds to the luminosity function of JWST-identified "Little Red Dots" reported by \citet{Kokorev2024}; the blue and orange curves represent the faint type-1 AGN population identified by \citet{Harikane2023} at two redshifts.
\textit{Right panel:} UV magnitude versus black hole mass relation for the disk-corona model (solid curve) and the broken power-law model (dash-dotted curve). Black hole masses are derived from the bolometric luminosities of each model, assuming Eddington-limited accretion in both cases and a radiative efficiency of 0.1 for the power-law model.
The scatter points show observed UV magnitude–black hole mass measurements compiled from \citet{Harikane2023}, \citet{Maiolino2024}, and \citet{Kokorev2024}. Black hole masses in the \citet{Harikane2023} and \citet{Maiolino2024} samples are estimated using H$\alpha$ line widths and luminosities. In the \citet{Kokorev2024} sample, black hole masses are inferred from bolometric luminosities - calculated assuming a radiative efficiency of 0.1 and Eddington-limited accretion - where the bolometric luminosities are estimated from the rest-frame 5100\,\AA~continuum luminosities obtained via SED fitting.
}
\label{fig:LF}
\end{figure*}

The AGN UV luminosity functions (UVLFs) adopted in this study are shown in the left panel of Fig.\,\ref{fig:LF}, highlighting substantial differences in their faint-end number densities. The solid curves represent UVLFs derived from the cumulative luminosity function of \citet{Jiang2022} and from the best-fit parameterizations of \citet{Harikane2023} and \citet{Kokorev2024}. It is evident that the type~1 AGN UVLF reported by \citet{Harikane2023} shows a significantly higher faint-end number density than those of \citet{Jiang2022} and \citet{Kokorev2024}.

The black curves in the right panel of Fig.\,\ref{fig:LF} show the monotonic relations between UV magnitude and BH mass predicted by the disk–corona model and by the broken power-law model, assuming Eddington-limit accretion. 
The scatter points represent observational estimates compiled from \citet{Harikane2023}, \citet{Maiolino2024}, and \citet{Kokorev2024}, where BH masses in the \citet{Harikane2023} and \citet{Maiolino2024} samples are estimated using H$\alpha$ line widths and luminosities. In the \citet{Kokorev2024} sample, black hole masses are inferred from bolometric luminosities - calculated assuming a radiative efficiency of 0.1 and Eddington-limited accretion - where the bolometric luminosities are estimated from the rest-frame 5100\,\AA~continuum luminosities obtained via SED fitting.
The figure shows that BH masses derived from our model exhibit better agreement with the observationally inferred values than those obtained using the broken power-law model. 
{Because BH mass, accretion rate, and UV luminosity are physically coupled in our model, any two of these quantities determine the third. 
In our calculations, we derive the BH mass from the observed UV luminosity assuming Eddington-limited accretion, using the resulting mass and accretion rate to construct the SED and estimate the corresponding number of ionizing photons.}

\section{The accretion rate of faint AGNs}\label{append:LRD_growth}

\begin{figure*}
\centering
{\includegraphics[width=2\columnwidth]{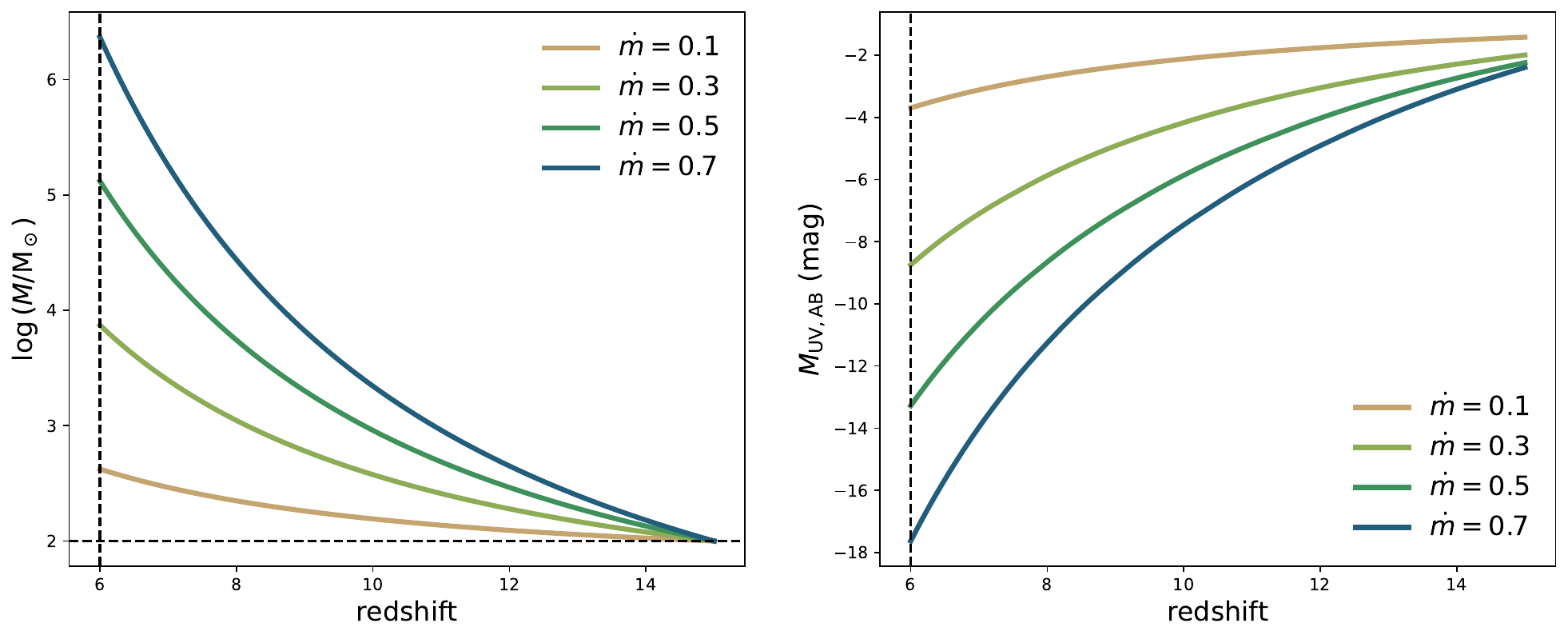}}
\caption{\textbf{Exponential growth of black hole mass and UV luminosities.} A black hole is seeded with an initial mass of $100\,\mathrm{M_\odot}$ at redshift $z_0 = 15$ and grows exponentially under constant Eddington-normalized accretion rates of $\dot{m} =$ 0.1, 0.3, 0.5, and 0.7. The vertical black dashed line marks redshift $z = 6$ for reference. The left panel shows the analytically computed BH mass growth histories as a function of redshift; the horizontal black dashed line indicates the initial seed mass. The right panel displays the corresponding absolute rest-frame UV magnitudes at 1450\AA. Analytical modeling of BH growth indicates that, in order to reach a mass of $\approx10^6 \mathrm{M_\odot}$ and $M_\mathrm{UV} \approx -18$ by $z = 6$, a minimum Eddington-normalized accretion rate of $\dot{m} \approx 0.7$ is required.
cross}
\label{fig:LRD_growth}
\end{figure*}

We estimate a lower limit on the accretion rate of faint AGNs using a simplified toy model. Specifically, we assume a black hole seeded at redshift $z = 15$ with an initial mass of $100\,\mathrm{M_\odot}$ - a typical value for a light SMBH seed widely adopted in the literature - accreting at a constant fraction of the Eddington-normalized rate $\dot{m}$. The black hole mass as a function of redshift, $M(z)$, is analytically given by $M(z)=M_0\,\mathrm{exp}\,[\dot{m}\times1.39\times10^{18}/\mathrm{M_\odot}\times(t(z)-t_0(z=15))]$, where $M_0 = 100\,\mathrm{M_\odot}$, $t(z)$ is the cosmic time at redshift $z$, and $t_0 = t(z=15)$. Fig.\,\ref{fig:LRD_growth} demonstrates the exponential growth of the black hole mass and the corresponding absolute rest-frame UV magnitude from $z = 15$ to $z = 6$ computed using the disk-corona model, for various values of $\dot{m}$ ranging from 0.1 to 0.7. In order to reach the characteristic mass and UV magnitude observed in LRDs, the accretion rate must be at least $\dot{m} \approx 0.7$.

\section{Secondary reionization of high-energy photons}\label{append:Xray_secondary}

\begin{figure}[h]
\centering
{\includegraphics[width=\columnwidth]{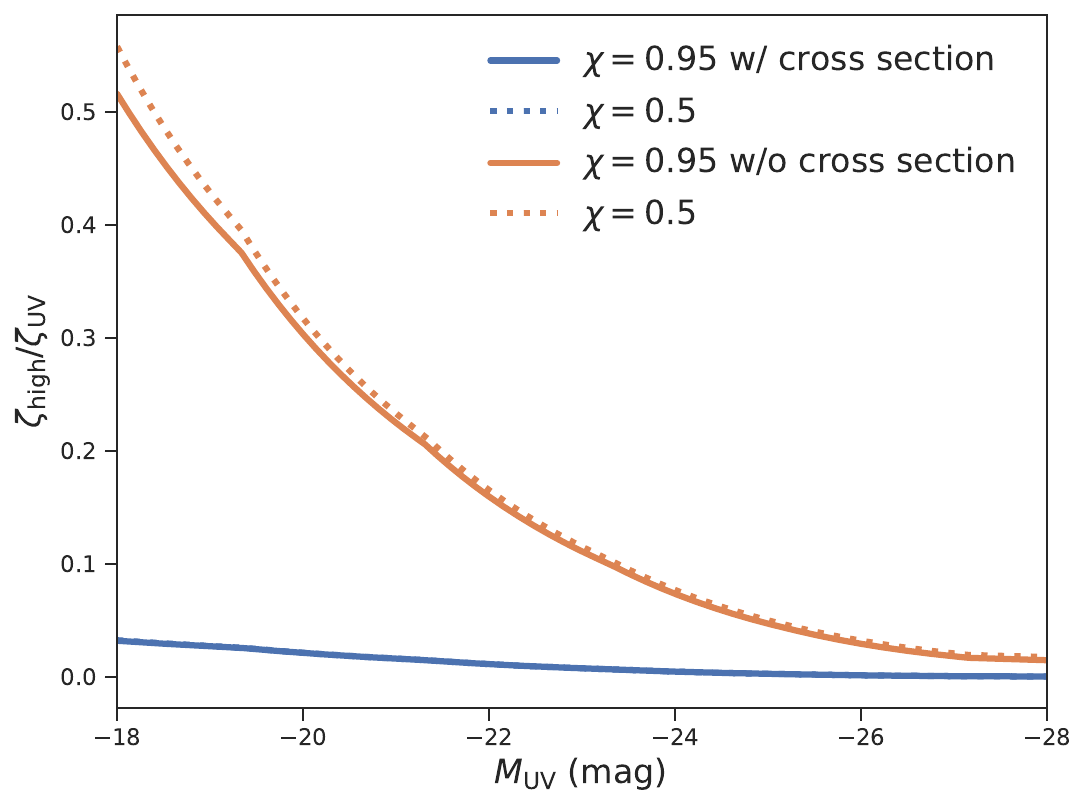}}
\caption{
{\bf The photoionization rate ratio produced by high-energy photons relative to UV photons.}
The relative influence of high-energy photon (4\,Ryd-10\,keV) ionization to that of UV photon (1-4\,Ryd) at each UV magnitude, considering both primary and secondary ionization events, where the solid and dotted curves show results corresponding to highly ionized ($\chi=0.95$) and mildly ionized ($\chi=0.5$) scenarios respectively.
The blue curves represent the photoionization rate ratios that consider the frequency-dependent cross section, and the red curves show the ratios ignoring the energy-dependence of the cross-section. 
}
\label{fig:multi_ionization}
\end{figure}

High-energy photons emitted by AGNs can, in principle, propagate over large distances and generate multiple secondary ionizations before becoming fully thermalized \citep{Shull1985, Madau2017}. The fraction of a photoelectron's energy that is deposited into secondary ionizations depends sensitively on the ionization state of the gas. To quantify this, we adopt the analytic fitting formula provided by \citet{Shull1985} for photoelectrons with energy $>100\,$keV,
\begin{equation}\label{eq:sec_frac}
f_\mathrm{sec} = 0.39[1-\chi^{0.41}]^{1.76},
\end{equation}
where $\chi$ is the ionization fraction, and $f_\mathrm{sec}$ is the energy fraction the photoelectron deposits into the ionization of secondary hydrogen atoms, reaching a maximum of $f_\mathrm{sec} = 0.39$ in nearly neutral gas ($\chi\approx0$). The end-of-reionization conditions discussed in the main text correspond to a highly ionized medium, for which $\chi = 0.95$ provides a reasonable approximation. For completeness, we also consider a moderately ionized scenario with $\chi = 0.5$ to examine the sensitivity to the ionization state. 

To estimate the number of hydrogen atoms ionized by a single photon, including both primary and secondary ionization events, we follow \citet{Shull1985, 2002ApJ...575...33R} and adopt the simplified expression
\begin{equation}
 N_\mathrm{a/p}= 1+f_\mathrm{sec}(E_\gamma-13.6\mathrm{eV})/13.6\,\mathrm{eV}    
\end{equation}
 where $E_\mathrm{\gamma}$ is the photon energy and $E_\gamma-13.6\,\mathrm{eV}$ is the kinetic energy of the resulting photoelectron. Here we define high-energy photons as those within the energy range $113.6\,\mathrm{eV}-10\,\mathrm{keV}$.
Because the resulting photoelectron carries relatively little excess energy in these regimes, UV photons (1–4 Ryd) produce essentially only a single ionization. For the same reason, photons with initial energies between 4 Ryd and 113.6 eV are likewise assumed to have $N_\mathrm{a/p}=1$. 

Applying this method to the fiducial accretion model $\dot{m}=1$, each observed UV magnitude is converted to a BH mass. Following \citet{2002ApJ...575...33R}, the photoionization rates of high-energy/UV photons can be calculated as, 
\begin{equation}\label{eq:photoion_rate}
\zeta = 4\pi\int_{\nu_0}^{\nu_1}N_\mathrm{a/p}\times\frac{J_\nu(m,\dot m)}{h\nu}\sigma_\nu\,\mathrm{d}\nu
\end{equation}
where $(\nu_0, \nu_1)=$(1, 4)\,Ryd for UV photons, and (4Ryd, 10keV) for high-energy photons; $J_\nu$ is the intensity of the central source (AGN); and $\sigma_\nu$ is the photon cross section, adopting the fitting formula given by \citet{1996ApJ...465..487V}, which sets the absorption probability that generally decreases with increasing photon energy. We neglect the effect of optical depth. We compute the ratio of photoionization rate 
$\frac{\zeta_\mathrm{high(4Ryd-10kev)}}{\zeta_\mathrm{UV}}$
and present the results in Fig.\,\ref{fig:multi_ionization}, where different linestyles denote results computed with different ionization fractions $\chi$, and different curve colors distringuish between calculations with and without the photon-energy-dependent cross section $\sigma_\nu$.

In the highly ionized case ($\chi = 0.95$), the ratio is consistently lower than that of the moderately ionized case ($\chi = 0.5$), but in both scenarios, the contribution from high-energy photons remains negligible (a few percent) once $\sigma_\nu$ is properly accounted for. If the cross-section dependence is ignored, the ratios increase to $\sim 52\%$ for $\chi = 0.95$ and $\sim 56\%$ for $\chi = 0.5$ at $M_{\rm UV} \approx -18$, and decline toward more luminous sources. These results indicate that ionizations driven by high-energy photons remain subdominant relative to those from UV photons across the full range of UV magnitudes considered.



\bibliography{sample701}{}
\bibliographystyle{aasjournal}



\end{document}